\documentclass
[  prl,
,  twocolumn   
,  amsfonts    
,  amssymb     
,  amsmath     
,  showpacs    
,  a4paper     
,  twoside     
] {revtex4-1}

\usepackage{color}
\usepackage{graphicx}
\usepackage{bm}
\usepackage{dcolumn}
\usepackage{verbatim}
\usepackage{import}

\definecolor{dark_red}{rgb}{0.75,0,0}
\definecolor{dark_blue}{rgb}{0,0,0.75}

\begin{document}
\bibliographystyle{prsty}

\title{Attosecond Transient Absorption Spectroscopy of doubly-excited states in helium}

\author{Luca Argenti$^1$}\email{luca.argenti@uam.es}
\author{Christian Ott$^2$}
\author{Thomas Pfeifer$^2$}
\author{Fernando Mart{\'i}n$^{1,3}$}
\affiliation{$^1$Departamento de Qu\'imica, M\'odulo 13, Universidad Aut\'onoma de Madrid, 28049 Madrid, Spain, EU}
\affiliation{$^2$Max-Planck Institut f\"ur Kernphysik, Saupfercheckweg 1, 69117 Heidelberg, Germany}
\affiliation{$^2$Center for Quantum Dynamics, Ruprecht-Karls-Universit\"at Heidelberg, 69120 Heidelberg, Germany}
\affiliation{$^3$Instituto Madrile\~no de Estudios Avanzados en Nanociencia (IMDEA-Nanociencia), Cantoblanco, 28049 Madrid, Spain}

\date{\today}

\begin{abstract}
Strong-field manipulation of autoionizing states is a crucial aspect of electronic quantum control. 
Recent measurements of the attosecond transient absorption spectrum of helium dressed by a few-cycle visible pulse [Ott et al., arXiv:1205.0519[physics.atom-ph] ] provide evidence of novel ultrafast resonant phenomena, namely, two-photon Rabi oscillations between doubly-excited states and the inversion of Fano profiles. 
Here we present the results of accurate \emph{ab-initio} calculations that agree with these observations and in addition predict that (i) inversion of Fano profiles is actually periodic in the coupling laser intensity and (ii) the supposedly dark $2p^2$ {$^1$S} state also appears in the spectrum. Closer inspection of the experimental data confirms the latter prediction.
\end{abstract}
\pacs{31.15.ac,\,\,32.80.Fb,\,\,32.80.Rm,\,\,32.80.Zb}

\maketitle

The fast development of attosecond laser pulses in the last decade 
gave access to the time-resolved study of correlated electron 
dynamics in atoms, molecules and solids on their natural time scale~\cite{Krausz2009}.
In this context, attosecond transient absorption spectroscopy 
(ATAS)~\cite{Goulielmakis2010} is emerging as a prominent technique 
complementary to the technologies which detect charged photofragments, 
like COLTRIMS~\cite{Ullrich2003a} and VMI~\cite{Eppink1997}, to monitor 
and control transiently bound states. Compared to photofragment detection,
ATAS provides higher energy resolution, and is applicable to condensed 
phases~\cite{Pfeifer2008} as well as to the gas phase, thus making it
a good candidate for the investigation of ultrafast electron dynamics 
in chemically relevant samples. Moreover, on the theoretical 
side, ATA spectra are significantly less demanding to reproduce than photoelectron 
distributions. While the latter require specification of scattering 
boundary conditions, the former only need the electronic wave function 
in the vicinity of the reaction center.

So far, ATAS has been employed to track both field-free ultrafast dynamics of 
coherent ensembles of valence-hole excited states~\cite{Goulielmakis2010,Wirth2011}, 
as well as the driven dynamics induced by strong external visible (VIS) 
or near-infrared (NIR) dressing fields~\cite{Wang2010e,Chini2012,Ott2012,
Ranitovic2011a,Holler2011,Shivaram2012}, including the following processes: 
electromagnetically induced transparency (EIT)~\cite{Ott2012,Ranitovic2011a};
instantaneous ac-Stark shift~\cite{Chini2012}; multiphoton transition matrix 
elements~\cite{Shivaram2012}; interference of wave-packets~\cite{Holler2011}; 
and time-resolved autoionization~\cite{Wang2010e}.

In a recent experiment~\cite{Ott2012}, ATAS has been used to investigate helium 
under the influence of a VIS dressing
field at tunable intensity in the energy region of the doubly-excited states converging to the N=2 
threshold. The experiment revealed clear Autler--Townes splitting of the $2s2p$ 
autoionizing state (the lowest term in the $sp_{2,2}^+$ {$^1$P$^o$} bright series)  
due to the coupling between this state, the $2p^2$~{$^1$S} state, and the 
consecutive $sp_{2,3}^+$ state, and inversion of the characteristic Fano profile 
asymmetry in the higher terms of 
the $sp_{2,n}^+$ series as a function of both the pump-probe time delay and of the 
intensity of the VIS dressing field.
\begin{figure}[hbpt!]
\centering
\includegraphics[scale=0.14]{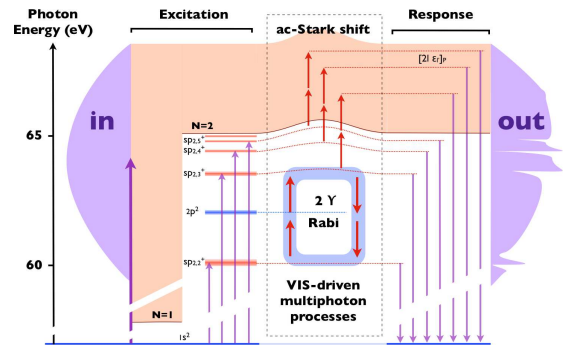}
\caption{\label{fig:1} {\bf Energy level scheme.} 
The attosecond transient absorption spectrum of helium records the 
linear response of the atom, dressed by a strong external VIS field, 
in the extreme ultraviolet (XUV) energy range. A sub-femtosecond broadband XUV 
pulse excites a wide range of states in the single-ionization region. 
The population and phase of the states so created is altered as a result 
of the interaction with the strong dressing field, due to multiphoton 
or even non-perturbative transitions. The XUV dipole response of the 
atom reflects such composition changes.
}
\end{figure}

Here we present the results of accurate \emph{ab-initio} calculations that reproduce 
the experimental conditions in~\cite{Ott2012}. We compute the TAS of a sub-femtosecond 
extreme ultraviolet (XUV) pulse for a helium atom dressed by a short (7\,fs) visible 
laser pulse (730\,nm), as a function of the time delay between the two pulses and of 
the peak intensity of the dressing field. 
Our calculations are in excellent agreement with most of the current experimental 
results and point towards further interesting quantum phenomena of the physics of 
two interacting electrons exposed to both the Coulomb field of the nucleus and a 
laser field. 
In particular, the inversion of Fano profiles is actually found to be periodic in 
the laser intensity, in agreement with the interpretation of this effect in terms 
of the energy shift of the doubly-excited states, which is approximately proportional
to the intensity of the dressing laser~\cite{Ott2012}.
Moreover, clear signatures of the 
supposedly dark $2p^2$ {$^1$S} state are observed. Finally, we show that the ATA
spectrum can be construed in terms of the 
non-diagonal component of the linear electrical susceptibility of the dressed atom.

In the ATAS experiment under examination, the weak 60~eV XUV pulse $\mathcal{E}(t)$ induces 
 a time-dependent dipole moment $d(t;\mathcal{E})$ in the atom dressed by the VIS field, which 
in turn results in a dipole emission that coherently contributes to the XUV field, thus altering 
its spectrum (see Fig.~\ref{fig:1}). 
For an optically thin medium, the relative change of the spectral intensity 
of the XUV pulse can be expressed in terms of an effective transient absorption atomic 
cross section $\sigma(\omega;\tilde{\mathcal{E}})$~\cite{Gaarde2011} 
\begin{equation}\label{eq:sigma_TAS_len}
\sigma(\omega;\tilde{\mathcal{E}})= \frac{4\pi\omega}{c}\,
\Im m\frac{\tilde{d}(\omega;\tilde{\mathcal{E}})}{\tilde{\mathcal{E}}(\omega)}
\end{equation}
where $\tilde{d}$ and $\tilde{\mathcal{E}}$ are the Fourier transform of the 
component of the dipole moment and the XUV field along the laser polarization, 
respectively, and where we emphasize the functional dependence of the response 
on the full spectrum of the impinging attosecond pulse.
In this process, strongly correlated doubly-excited states (DES) have a preponderant role. 
For the latter, common models to evaluate the rate of non-perturbative processes, 
like ionization by tunneling~\cite{Ammosov1986,Perelomov1966}, are inapplicable, due to the
complete breakdown of the single-active-electron (SAE), the adiabatic, and the non-polarizable
parent-ion approximations~\cite{Smits2004,Pfeiffer2011}.
In this context, calculations based on simplified 
models~\cite{Lambropoulos1981,Bachau1986,Loh2008,Themelis2004,Chu2011,Chu2012,Chu2012a,Chini2012}, 
SAE, or stationary methods, like 
Floquet theory~\cite{Gaarde2011,Tong2010b} have therefore limited applicability. 
The interpretation of these experiments require instead both a complete \emph{ab-initio} 
representation of the system and a direct solution of the time-dependent 
Schr{\"o}dinger equation (TDSE).

Here, we solve the TDSE with a Krylov representation of a second-order exponential unitary 
time-step propagator in velocity gauge~\cite{Argenti2010}. Reflection from the box 
boundaries is prevented by complex absorbing potentials. The wave function is expanded 
on a two-particle spherical basis, the angular part is represented by bipolar spherical 
harmonics~\cite{Varshalovich1988} and the radial part by B-splines with an asymptotic 
spacing of $0.5$~au. Each total angular momentum comprises all the partial-wave channels 
with configurations of the form $N\ell\epsilon_{\ell'}$ with $N\leq 2$, and a full-CI 
localized channel $n\ell n'\ell'$ that reproduces short-range correlations between the 
two electrons~\cite{Argenti2006}. Such representations of the correlated electron basis 
is able to represent in a concise way and with high accuracy all the resonant series in 
helium converging to any given threshold $N$~\cite{Argenti2006},
and, thanks to the localized channel, the first few doubly-excited states converging to 
thresholds not included in the close-coupling expansion as well.
The ATAS response is evaluated with a velocity-gauge analogue of~(\ref{eq:sigma_TAS_len}), 
obtained via an application of the Ehrenfest theorem
\begin{equation}\label{eq:sigma_TAS_vel}
\sigma(\omega;\tilde{\mathcal{E}})=
-\frac{4\pi}{\omega}\,\Im m\,\frac{\tilde{p}(\omega;\tilde{A})}{\tilde{A}(\omega)},
\end{equation}
where $\tilde{A}$ and $\tilde{p}$ are the FT of the vector potential and the canonical 
momentum. The momentum $\tilde{p}$ is given by the sum of two complementary terms: 
$\tilde{p}_-$, computed numerically while in the presence of the dressing field, and 
$\tilde{p}_+$, computed analytically for the subsequent field-free evolution. The result
is thus equivalent to the one that would be obtained for a simulation protracted for an
infinite time.
In the simulation, we include the states with orbital angular momentum up to $\ell_{max}=5$ 
for the localized channel, and a total angular momentum up to $L_{max}=10$. We checked for
convergence with respect to both these limits by repeating few calculations with either 
$L_{max}=15$ or $\ell_{max}=10$, as well as by including the $N=3$ partial-wave channels, 
without observing any significant change in the result.

\begin{figure}[hbpt!]
\centering
\includegraphics[scale=0.23]{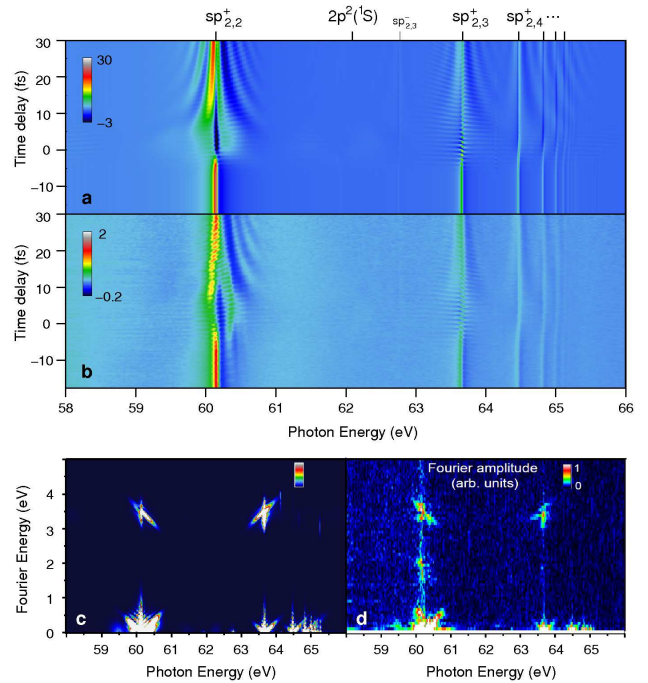}
\caption{\label{fig:2} ATAS as a function of the time delay.
a) Theoretical transient absorption spectrum as a function of the 
XUV photon energy (x-axis) and of the time delay (y-axis) between the XUV
and the VIS pulses, for a VIS intensity I=3.5~TW/cm$^2$. b) Matching 
experimental spectrum recorded for a nominal VIS intensity of 2.5$\pm$1.5~TW/cm$^2$.
c) Theoretical, d) Experimental Fourier transform with respect to the time delay of the spectra shown
in panel a-b).
}
\end{figure}
In Fig.~\ref{fig:2}-a,b we compare the present theoretical ATAS spectrum $\sigma(\omega;\tau)$,
as a function of the photon energy and the time delay $\tau$, to the experimental optical density 
(OD)~\cite{Ott2012} (OD~$\propto\sigma$).
The agreement is excellent. The vertical features visible in the two spectra are due to the bright 
$sp^+_{2,n}$ series of doubly-excited states, $n=2,\,3,\,\ldots$~\cite{Cooper1963}. At large negative time delays, when the dressing pulse comes well before the XUV 
pulse, the ATAS spectrum displays the characteristic Fano profiles~\cite{Fano1961,Madden1963,Cooper1963} 
of the field-free atom. When the two pulses overlap, the $sp_{2,2}^+$ state separates in two branches; this is
 the well known~\cite{Bachau1986,Chu2012,Ott2012} Autler--Townes splitting (ATS) associated with 
the strong coupling between the $sp_{2,2}^+$ and the $2p^2$~{$^1$S} state, which, with the current energy of the dressing 
field, 1.7~eV, is resonantly coupled to both the first and the second term of the $sp_{2,n}^+$ 
series (see Fig.~\ref{fig:1}). 
The faint transversal wavelike features visible around both the $sp_{2,2}^+$ and the $sp_{2,3}^+$ terms for
positive time delays are slices of the hyperbolic branches $\tau_n(E)=2\pi n/|E-E_0|$. 
These features
are caused by the beating between the amplitude of the resonantly-enhanced two-photon transition from one XUV-excited 
autoionizing state to the continuum-coupled other resonant state and the amplitude of the XUV transition from the ground state to 
the same coupled-resonant and continuum states. As far as the direct-ionization part is concerned, such 
phenomenon is the transient-absorption analogue to the one described for the photoelectron spectrum 
in~\cite{Mauritsson2010}.
At very large time delays, the signal converges again to the field-free case. Here, the fringes
converging around each resonance are a consequence of the sharp change in the phase or population 
of the localized part of the resonances induced by the strong VIS pulse.
Finally, in the dark narrow region between the two Autler--Townes branches of the $sp_{2,2}^+$ state 
the effective transient absorption cross section is \emph{negative}, what indicates that
the XUV field in that region is actually amplified rather than absorbed, as a result of the
interaction with the atom. This observation confirms previous predictions~\cite{Schapper2010,Gaarde2011,Chu2012} 
and experimental findings~\cite{Wirth2011}. 

In Fig.~\ref{fig:2}-c,d we compare the FT with respect to the time delay of the ATA spectra in 
Fig.~\ref{fig:2}-a,b~\footnote{For the sake of clarity, we have removed the spectral asymptote $\sigma(\omega,\infty)$, which corresponds to the optical response of the field-free atom.}
\begin{equation}\label{eq:DefFTSigma}
\tilde{\sigma}(\omega,\omega_\tau)=\int_{-\infty}^\infty d\tau\,\frac{e^{-i\omega_\tau\,\tau}}{\sqrt{2\pi}}
\left[\sigma(\omega,\tau)-\sigma(\omega,\infty)\right],
\end{equation}
where $\omega_\tau$ is the Fourier Energy.
To interpret such bi-dimensional spectra, let us consider again Eq.~\ref{eq:sigma_TAS_len}.
Since the XUV pulse is weak, the dipole response of the laser-driven system can be 
assumed to be linear
\begin{equation}
\tilde{d}(\omega;\mathcal{E})\,=\,\int d\omega' \chi(\omega,\omega')\tilde{\mathcal{E}}(\omega').
\end{equation}
where $\chi(\omega,\omega')$ is the electric susceptibility of the dressed atom. 
\begin{figure}[hbpt!]
\centering
\includegraphics[scale=0.22]{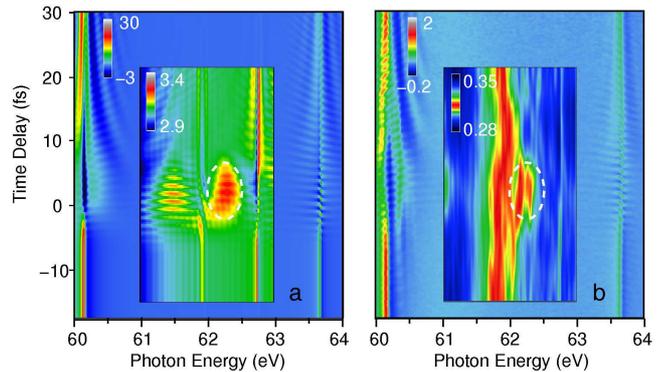}
\caption{\label{fig:2p2} 
Detail of the comparison between theoretical (left panel) and experimental 
(right panel) ATAS as a function of the XUV photon energy 
(x-axis) and of the time delay between VIS and XUV pulses (y-axis) in the
energy region between the $sp_{2,2}^+$ and the $sp_{2,3}^+$ doubly-excited states.
In the insets, closer inspection unveils a clear signature of the 
$2p^2$ {$^1$S} resonance, which mediates the transition between the lowest 
$^1$P bright states.}
\end{figure}
For time-invariant samples, of course, the susceptibility is diagonal: $\chi(\omega,\omega')=\delta(\omega-\omega')\chi_d(\omega)$. This is not the case here, however. Under the action of 
any given excitation frequency, a dressed atom will in general give rise to a response 
at other frequencies as well. This means that, alongside the field-free diagonal component, 
the electrical susceptibility also comprises a non-diagonal component $\chi_{nd}$,
\begin{equation}
\chi(\omega,\omega')=\delta(\omega-\omega')\chi_d(\omega)+\chi_{nd}(\omega,\omega').
\end{equation}
For moderate intensities of the dressing lasers, the strongest features of $\chi_{nd}$ 
are expected to be located close to the diagonals $\omega=\omega'\pm 2n\omega_{vis}$.
If the spectrum of the exciting laser is narrower than $2\omega_{vis}$, as 
in~\cite{Buth2007,Buth2008,Gaarde2011}, transient absorption spectroscopy only probes 
the response close to the central diagonal $\omega=\omega'$. This provides valuable 
yet incomplete information on the ultrafast optical response of the dressed system.
In the present ATAS, on the other hand, the spectral width of attosecond pulse exceeds 
$2\omega_{vis}$, therefore the response at $\omega=\omega'\pm 2\omega_{vis}$ results 
in a detectable heterodyne signal. This result holds for isolated as well as for
trains~\cite{Ranitovic2011a,Holler2011,Shivaram2012} of attosecond pulses.
The variation of the XUV field due to the change in time delay $\tau$ with respect 
to the dressing field, which defines the time origin, is readily parametrized as
\begin{equation}\label{eq:td_dependence_of_E}
\tilde{\mathcal{E}}(\omega)\,=\tilde{\mathcal{E}}(\omega;\tau)\,=\,e^{-i\omega \tau}\,\tilde{\mathcal{E}}_0(\omega),
\end{equation}
where $\tilde{\mathcal{E}}_0$ corresponds to a pulse centered at the time origin.
From Eqns.~(\ref{eq:sigma_TAS_len}-\ref{eq:td_dependence_of_E}) and in the limit 
of extremely short single pulses, where $\tilde{\mathcal{E}}_0(\omega)$ is 
eventually constant in an interval of the order of the excitation frequencies of 
the system, we obtain
\begin{equation}
\frac{ic\,\,\tilde{\sigma}(\omega,\omega_\tau)}{(2\pi)^{3/2}\omega}\,\,\,\simeq\,\,\,
\chi_{nd}(\omega,\omega-\omega_\tau)\,-\,\chi_{nd}^*(\omega,\omega+\omega_\tau)
\end{equation}
The strong signals close to the $\omega_\tau=0$ axis, therefore, are associated
to the electromagnetically induced transparency (EIT) of the DES that is
visible also in femtosecond TAS~\cite{Loh2008,Gaarde2011,Chu2012a}. The two 
strong features above the $sp_{2,2}^+$ and the $sp_{2,3}^+$ resonances at
$\omega_\tau=2\omega_{vis}=3.4$~eV, instead, are heterodyne signals that
are visible only in ATAS. In particular, the one above the $sp_{2,2}^+$ state
can be assigned to the  $\chi_{nd}^*(\omega_{sp_{2,3}^+}-2\omega_{vis},\omega_{sp_{2,3}^+})$
component of $\tilde{\sigma}$, while the one above the $sp_{2,3}^+$ state 
to the  $\chi_{nd}(\omega_{sp_{2,2}^+}+2\omega_{vis},\omega_{sp_{2,2}^+})$ component.
Both signals are strongly enhanced by the presence of the resonant $2p^2$~{$^1$S} 
intermediate state (compared to the higher terms in the $sp_{2,n}^+$ series) and 
comprise a non-resonant as well as a resonant part. The latter is indicative
of an incipient resonant two-photon Rabi oscillation between the two {$^1$P$^o$}
states (see Fig.~\ref{fig:1}).

As a consequence of its even parity, the $2p^2$~{$^1$S} is normally a ``dark'' 
state. Yet, due to the effect of the dressing field both on it and on the ground 
state, the dipole amplitude between these two states is not exactly zero. 
Indeed, two faint signals at photon energy $\sim 61.5$~eV and $~\sim 62.3$~eV 
close to the $\omega_\tau=0$ axis are visible in Fig.~\ref{fig:2}.
To confirm this observation, in figure~\ref{fig:2p2} we show the theoretical 
and experimental ATAS with enhanced-contrast insets close to where the $2p^2$~{$^1$S} 
is expected to emerge. The upper Autler-Townes branch of this state is clearly 
visible in both insets with an intensity comparable to that of the narrow 
$sp_{2,3}^-$ {$^1$P$^o$} resonance. This result shows that TAS is not restricted to 
the study of bright states only. Also ``dark'' states which are coupled to 
bright ones can emerge.

\begin{figure}[hbpt!]
\centering
\includegraphics[scale=0.36]{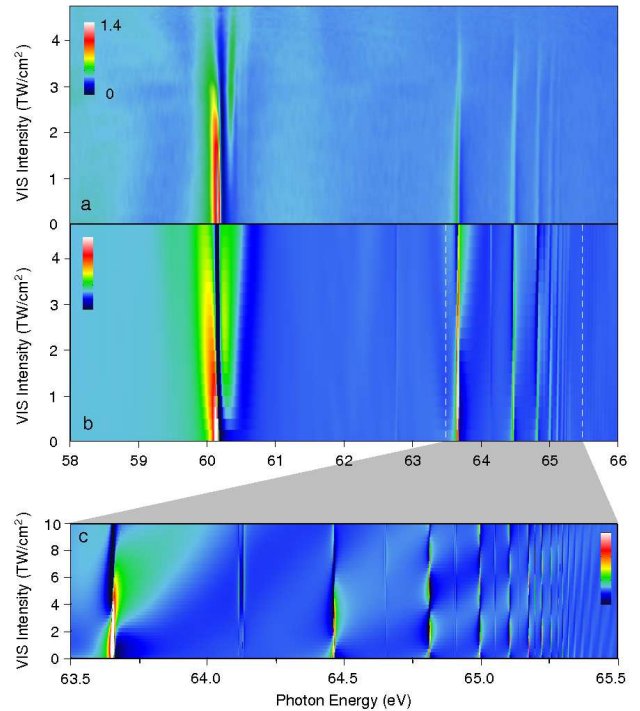}
\caption{\label{fig:ATAS_vs_EI} 
ATAS for overlapping XUV and VIS pulses as a function of the XUV photon energy 
(x-axis) and of the intensity of the VIS dressing field (y-axis): a) 
experimental results (OD); b) theoretical prediction; c) detail of the theoretical 
prediction (notice the extended intensity range).
The experimental and theoretical results are in good qualitative agreement 
at low intensities. While the experimental signal seems to fade at 
I=4TW/cm$^2$, the calculations predict ongoing resonant features
up to intensities as high as 10~TW/cm$^2$. In particular, the characteristic
flip in the asymmetry of the resonant Fano profiles doesn't occur just
once, as observed in the experiment; instead, it is periodic in the 
intensity of the dressing field. See text for more details.
}
\end{figure}
Fig.~\ref{fig:ATAS_vs_EI}a shows the experimental ATAS for overlapping XUV and VIS
pulses, as a function of the intensity of the dressing laser. The spectrum is 
averaged over a time-delay window of 1.2~fs to reduce the noise. Fig.~\ref{fig:ATAS_vs_EI}bc
show the theoretical prediction for a fixed time delay $\tau=200$~au (maximum ATS).
Overall, except for the highest intensities, the agreement is good:
we observe the increase of the ATS of the $sp_{2,2}^+$ resonance as well as the
inversion of the asymmetry of the Fano profile for all the higher terms in 
the $sp_{2,n}^+$ series as the intensity of the dressing field increases. 
In~\cite{Ott2012}, this latter effect has been attributed to the 
ac-Stark shift $\Delta E_{ac-Stark}(t)$ of the localized component autoionizing 
states. The energy shift translates into an extra phase 
$\Delta\phi(\tau)=-\int_{\tau}^\infty \Delta E_{ac-Stark}(t')dt'$ between resonant 
and background amplitude that is approximately proportional to the peak intensity 
of the laser. As a consequence, the asymmetry (Fano profile) inversion is expected 
to be periodic in the laser intensity. Our calculations (see Fig.~\ref{fig:ATAS_vs_EI}) 
clearly confirm this behavior. In the experimental spectrum, 
all resonant features suddenly fade for intensities larger than 4~TW/cm$^2$, 
close to the instrumental limit, which is sufficient to observe just the 
first inversion for the highest terms in the series.  
Convolution of the theoretical results in Fig.~\ref{fig:ATAS_vs_EI}c with the 
experimental energy and intensity resolutions ($\sigma_E\simeq$20~meV and 
$\sigma_I\simeq 20\%\,I$, respectively) is found to smear out the resonant
features at intensities higher than $5$~TW/cm$^2$. 
The ac-Stark shift of singly-excited states in helium are known to approach 
the ponderomotive energy~\cite{Tong2010b}, which is the limiting value expected 
for a Rydberg electron bound to an unpolarizable core~\cite{Davidson1993}. 
For the highest terms of the autoionizing $sp_{2,n}^+$ series, instead, our calculations 
show that the ac-Stark shift is significantly larger than such limit.
In fact, in DES the two electrons are strongly correlated: for the higher terms in 
the $sp_{2,n}^+$ series, e.g., the parent ion is permanently polarized towards the outer 
electron. This is further indication that traditional formulas for estimating 
polarizability, tunneling rates and ac-Stark shifts are inapplicable to DES.
At higher intensities, not shown here, we observe a multi-branch splitting 
of the $sp_{2,2}^+$ doubly-excited state that we associate to the multi-photon 
Rabi oscillation between the  $sp_{2,2}^+$, the $2p^2$ {$^1$S} and the $sp_{2,3}^+$ 
states.

Some simplifying assumptions in the calculations are likely responsible for the 
occasional discrepancies between theory and experiment. In particular, the 
theoretical XUV pulse has a purely Gaussian spectrum, while the experimental 
spectrum shows some residual odd-harmonic modulation. Furthermore, we do not 
represent the wide pedestal that is known to accompany the main VIS dressing 
pulse in practice. Finally, we didn't take advantage of the systematic relative 
uncertainty, of the order of $\pm20\%$, introduced by the calibration of the 
VIS intensity absolute value. Even so, the remarkable agreement with either the 
data recorded at the nominal intensity of $3.5$~TW/cm$^2$ reported in~\cite{Ott2012}
and those used in the present work is evidence that our conclusions are robust 
and sound.

In conclusion, in this work we showed how the inversion of asymmetric resonant
transient absorption profiles in ATAS of dressed helium is periodic 
in the intensity of the driving laser, and that the field-free dark
$2p^2$~{$^1$S} state actually gives rise to a distinct signal in
the ATA spectrum under the influence of a moderately strong VIS laser field.
The relation of TAS spectra to the non-diagonal component of the 
electrical susceptibility highlighted in this work offers an 
alternative and rational perspective on ATAS spectroscopy and 
suggests a direct way to measure and thus modulate the ac-Stark 
shift in transiently bound states, which is an important tool for 
the implementation of quantum-control protocols.
Finally, the observation that the ac-Stark shift of autoionizing states 
close to the N=2 threshold is larger than the theoretical limit in 
the single-active electron approximation is an indication that 
common model tunneling-rate formulas are inapplicable to such states
and, by extension, to any multi-electron atom where polarization of
the core affects the effective field strength experienced by the 
ionizing active electron.

We thank Eva Lindroth for useful discussions during the early stages of 
this investigation. We thank Mare Nostrum BSC and CCC-UAM (Centro de 
Computaci\'on Cient\'ifica, Universidad Aut\'onoma de Madrid) for 
allocation of computer time.
The research leading to these results has received funding 
from the European Research Council under the European Union's 
Seventh Framework Programme (FP7/2007-2013)/ERC grant agreement 
n290853, the European COST Action CM0702, the ERA-Chemistry project 
n$^\circ$ PIM2010EEC-00751, the Marie Curie ITN CORINF, the MICINN 
project n$^\circ$ FIS2010-15127, ACI2008-0777 and CSD 2007- 00010 (Spain).
CO and TP acknowledge financial support by the Max-Planck 
Research Group Program of the Max-Planck Gesellschaft (MPG).

\end{document}